# Response to "Room Temperature, Quantum-Limited THz Heterodyne Detection? Not Yet"

Mona Jarrahi and Yen-Ju Lin

*Electrical and Computer Engineering, University of California – Los Angeles, CA, 90095, USA*

*Correspondence to Mona Jarrahi, [mjarrahi@ucla.edu](mjarrahi@ucla.edu)*

This commentary is written in response to arXiv:1907.13198. In this article, Zmuidzinas et al. raise questions about the results reported by our group in *Nature Astronomy* [1], DOI: 10.1038/s41550-019-0828-6, regarding our experimental methodology and our device performance metrics. As described below, Zmuidzinas et al. have unfortunately missed some basic principles on impedance matching and the physics of photomixers and plasmonics that are at the heart of their categorical conclusions. Here, we correct their misunderstandings and discharge all of their flawed conclusions as detailed below. All of the results and conclusions reported in our *Nature Astronomy* manuscript remain correct, as before.

A. Impedance Matching

Zmuidzinas et al. mistakenly estimate a severe impedance mismatch between our photomixer and backend IF electronics and come to the false conclusion that:

'…*The very high value reported for their photomixer impedance strongly suggests that the conversion loss is quite poor and that the actual sensitivity is far worse than claimed.*'

'…*Their theory is therefore incompatible with their claimed sensitivity, which requires a large internal conversion gain to overcome the severe IF impedance mismatch as discussed above.*'

'…*Judging from the large photomixer impedance reported by Wang et al., the sensitivity is likely ∼100× worse than claimed.*'

Surprisingly, for their impedance matching assessment, Zmuidzinas et al. assume that the IF impedance seen by the 25 kΩ photomixer at 1 GHz is 50 Ω, while totally ignoring the fact that the IF current induced at the photomixer active area (with micrometer scale dimensions) should pass through the logarithmic spiral antenna, bond pads, bond wires, and transmission lines to reach the SMA connector (with millimeter scale dimensions) that routs the IF signal to the backend IF electronics and the fact that these components all have impacts on the impedance seen by the photomixer.. Unfortunately, Zmuidzinas et al. have missed this important fact and came to the wrong conclusion, which is summarized in their Letter as:

'*There is a severe impedance mismatch between the 25kΩ photomixer and the 50Ω IF amplifier, corresponding to a coupling loss of 21 dB. For the overall conversion loss to be no worse than 3 dB, as required by the reported sensitivity, the photomixer would need to have an internal conversion gain of at least 18 dB!*'

In fact, in contrary to their incorrect assessment, the IF coupling loss from photomixer to backend IF electronics is ~4.5 dB, much lower than the 21 dB value estimated by Zmuidzinas et al. This is because the logarithmic spiral antenna, bond pad, bond wire, and printed circuit board that connect the photomixer to the backend IF electronics (Fig. 1a) offer an impedance transformation from 50 Ω to 2697+j486 Ω at 1 GHz. More details are given below.

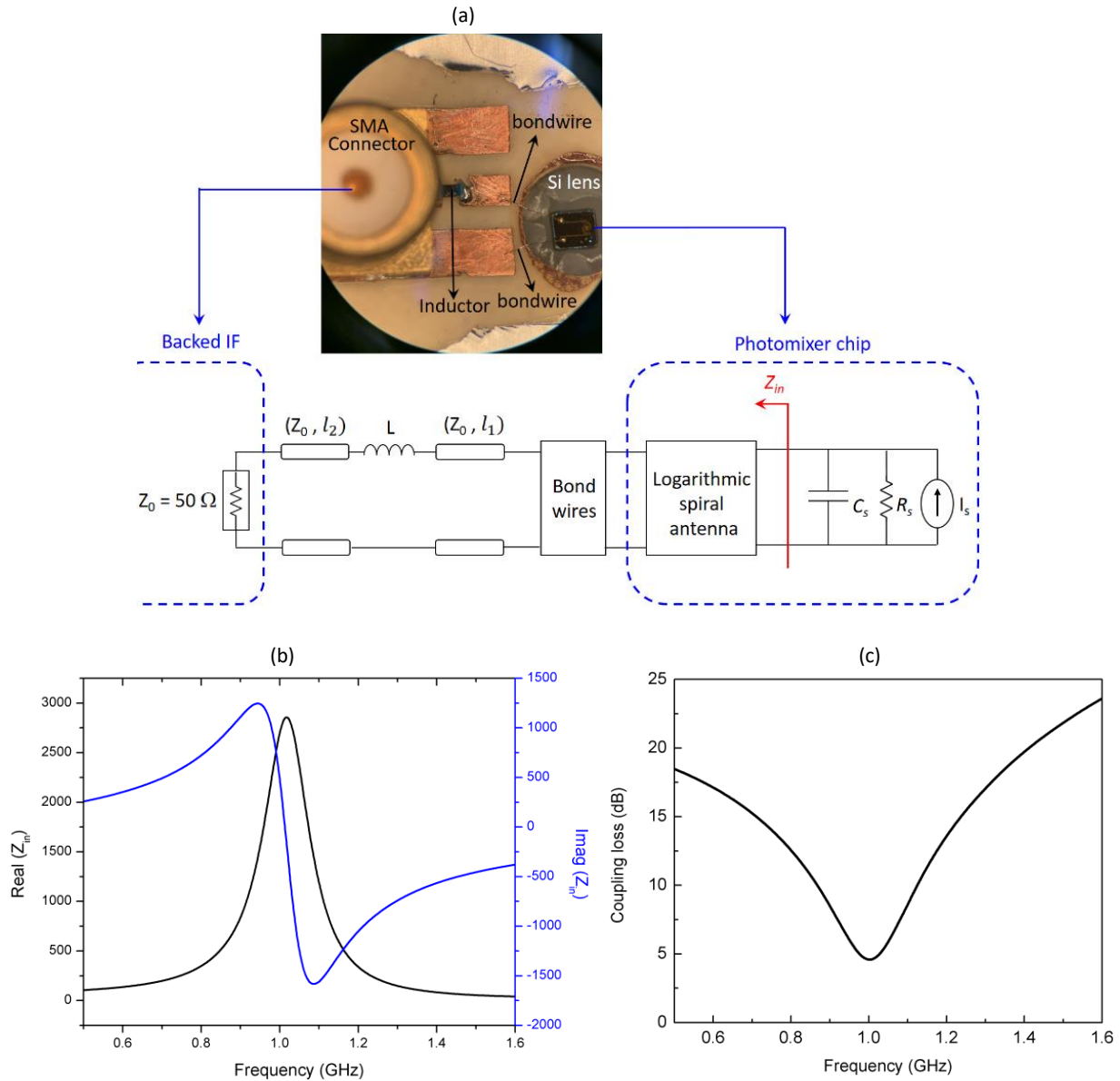

Figure 1. (a) Image of the printed circuit board connecting the plasmonic photomixer to the backend IF electronics and its equivalent circuit model. The estimated impedance observed by the photomixer active area and the resulting IF coupling loss using ADS simulations are shown in (b) and (c), respectively.

Figure 1a shows the image of the printed circuit board (a 1.6 mm-thick Roger RO4003C substrate) that connects our photomixer to the SMA connector, which routes the IF signal to the backend, and its equivalent circuit model. The photomixer active area is modeled as a current source in parallel with the photomixer resistor ($R_s$ ~ 25 kΩ) and capacitor ($C_s$ ~ 1 fF). ADS simulations are used to estimate the input impedance observed by the photomixer active area, $Z_{in}$. For this simulation, the frequency response of the logarithmic spiral antenna including the bond pads and the bond wires are first estimated using HFSS simulations and exported to ADS. The geometry of the coplanar transmission line that is fabricated on the printed circuit board is selected to achieve a 50 Ω characteristic impedance. An SMD inductor (L = 56 nH) is placed in series with the transmission line. The length of the coplanar transmission line on each side of the inductor is $l_1$ = 3.4 mm and $l_2$ = 6.25 mm. The estimated input impedance observed by the photomixer active area is shown in Fig. 1b, suggesting a $Z_{in} = 2697 + j486.2$ Ω at 1 GHz and a coupling loss of 4.5 dB. While <u>direct</u> measurement of this coupling loss is not possible due to the micrometer scale dimensions of the photomixer active area, one-port scattering parameter measurements ($S_{11}$) from the SMA connector confirm the accuracy of the circuit model used for estimating the coupling loss.

B. Photomixer Device Physics and Basics

Zmuidzinas et al., have a rather unfortunate and surprisingly big misunderstanding of the operation principles of our plasmonic photomixer, which has led to many of the incorrect assumptions they have made in their statements. Despite the detailed description of the device physics in our manuscript and supplementary Fig. S2, which clearly explain the role of the optical pump beam and plasmonic enhancement in boosting the THz-to-RF conversion efficiency, Zmuidzinas et al. state that:

'*…Indeed, the theory of operation presented by Wang et al. makes no mention of conversion gain. According to equation (6) of their supplemental information, they assume a linear, local relationship between the current density and THz electric field at frequency $f_{THz}$. For $f_{THz}$ < 1/2πτ ≈ 500 GHz where τ = 0.3ps is the stated carrier lifetime, the response is essentially instantaneous and their equation (6) is equivalent to Ohm's law J(r,t) = σ(r,t) E(r,t). Here σ(r,t) is the (real) conductivity that varies with time t and position r due to the photogeneration of carriers by the two lasers with beat frequency $f_{beat}$. The equation J = σ E simply describes current flow in an ordinary resistor, and leads to the circuit version of Ohm's law I = V/R, or I(t) = V (t)/R(t) when the conductivity is time dependent. Here I(t) and V (t) are the current and voltage across the terminals of the THz spiral antenna, and the R(t) is the time-dependent photomixer resistance as seen from the antenna terminals. Thus, the theory offered by Wang et al. places the device into a well-known class of "resistive mixers" that includes diode mixers and FET mixers, which are not capable of conversion gain and in fact are subject to a theoretical minimum conversion loss of 3 dB.*'

The surprising argument that the equation (6) of the supplementary information resembles Ohm's law, placing the photomixer into the class of "*resistive mixers*" such as diode and FET mixers is a categorical mistake. In their incomplete circuit model of the photomixer, Zmuidzinas et al. miss to include the important current source component (Fig. 1a) that is proportional to the optical pump power and can be significantly boosted by plasmonic enhancement effects, which is the source of conversion gain in the THz-to-RF conversion process by injecting photocarriers to the system.

This flawed understanding of the operation principles of our plasmonic photomixer has further misled Zmuidzinas et al. in distinguishing the major differences between our plasmonic photomixer and other mixers used in conventional heterodyne receivers. For example, despite the detailed explanation in the supplementary Fig. S2 describing the governing device operation equations, Zmuidzinas et al. assume that our plasmonic photomixer is capable of direct hot/cold radiation detection in the absence of the optical pump beam similar to diode and FET mixers that are capable of direct detection of the hot/cold radiation in the absence of a local oscillator. Based on this wrong assumption, Zmuidzinas et al. also question the validity of our Y-factor measurements, which this is completely unjustified as detailed below.

C. Y-factor Measurements

Zmuidzinas et al. make two major mistakes when commenting on our Y-factor measurement technique, which bring them to the conclusion of '…*fatally flawed sensitivity measurements*'.

Their first mistake involves misunderstanding of lock-in amplifier capabilities. Zmuidzinas et al. are wrong when assuming that lock-in amplifier can only measure changes (difference) in the IF output that occur in response to the modulation of the optical pump. In fact, one can also measure and record the time-domain variations in the detected signal by a lock-in amplifier. For our measurements, we used the scope module of our lock-in amplifier (Zurich Instruments MFLI) to monitor the IF output both when the optical pump beam was ON and OFF for hot and cold loads ($P_{IF,hot,on}$ and $P_{IF,hot,off}$ for the hot load and $P_{IF,cold,on}$ and $P_{IF,cold,off}$ for the cold load). Therefore, unlike Zmuidzinas et al.'s assumption, the actual IF power values can be extracted from the power detector's calibration curve and the addition of a constant value to all the power values would not have gone unnoticed in our experiments.

Their second mistake involves their wrong understanding of the photomixer device physics and failing to differentiate it from that of other mixers used in conventional heterodyne terahertz detectors, as they incorrectly say '*the theory offered by Wang et al. places the device into a well-known class of "resistive mixers" that includes diode mixers and FET mixers, which are not capable of conversion gain and in fact are subject to a theoretical minimum conversion loss of 3 dB*'. By this incorrect understanding, Zmuidzinas et al. assume that our photomixer is capable of direct detection of the hot/cold radiation in the absence of the optical pump beam, similar to diode mixers and FET mixers that are capable of direct detection of the hot/cold radiation in the absence of a local oscillator. Therefore, they inaccurately state that the values of $P_{IF,hot,off}$ and $P_{IF,cold,off}$ should be used to make corrections to the Y-factor values:

'*The paper makes no mention of measurements of $P_{IF,hot,off}$ or $P_{IF,cold,off}$, and because this information is missing, it is not possible to convert the reported values of Y ' into corrected values for Y . Noise temperatures calculated using Y ' are meaningless, and can be radically different from the true noise temperatures calculated using Y. Thus, one cannot place any confidence in the sensitivities reported by Wang et al. in their Figure 3c.*'

Unfortunately, Zmuidzinas et al. miss the fact that $P_{IF,hot,off}$ and $P_{IF,cold,off}$, which have the same value, are simply the background noise of the device when the optical pump beam is turned OFF (i.e., photomixer is turned OFF) and are independent of the radiation received from the hot and cold loads and that $P_{IF,hot,on} - P_{IF,hot,off}$ and $P_{IF,cold,on} - P_{IF,cold,off}$ represent $P_{IF,hot,on}$ and $P_{IF,cold,on}$ values corrected for common mode noise that dominate the $P_{IF,hot,off}$ and $P_{IF,cold,off}$ values, respectively. Lock-in detection is an obvious way to cancel out the impact of the common mode noise in a detection system. It has been extensively used for Y-factor measurements of conventional heterodyne terahertz detectors and its effectiveness in accurate assessment of the Y-factor and noise temperature compared to the techniques that do not use lock-in detection has been already demonstrated [2]. While lock-in detection is a legitimate way to eliminate some of the common mode background noise in our detection system, it is worth mentioning that the differences between the Y-factor values calculated with lock-in detection $(P_{IF,hot,on} - P_{IF,hot,off})/(P_{IF,cold,on} - P_{IF,cold,off})$ and without lock-in detection $P_{IF,hot,on}/P_{IF,cold,on}$ (Zmuidzinas et al.'s recommended approach) for our measurement data are much lower than the fluctuations in the Y-factor values due to the fluctuations in the optical pump power (error bars shown in Fig. 3a of the manuscript). Similarly, the extracted noise temperature values from the Y-factor data calculated with lock-in detection $(P_{IF,hot,on} - P_{IF,hot,off})/(P_{IF,cold,on} - P_{IF,cold,off})$ and without lock-in detection $P_{IF,hot,on}/P_{IF,cold,on}$ (Zmuidzinas et al.'s recommended approach) are much lower than the fluctuations in the noise temperature values due to the fluctuations in the optical pump power (error bars shown in Fig. 3c of the manuscript), as shown in Fig. 2.

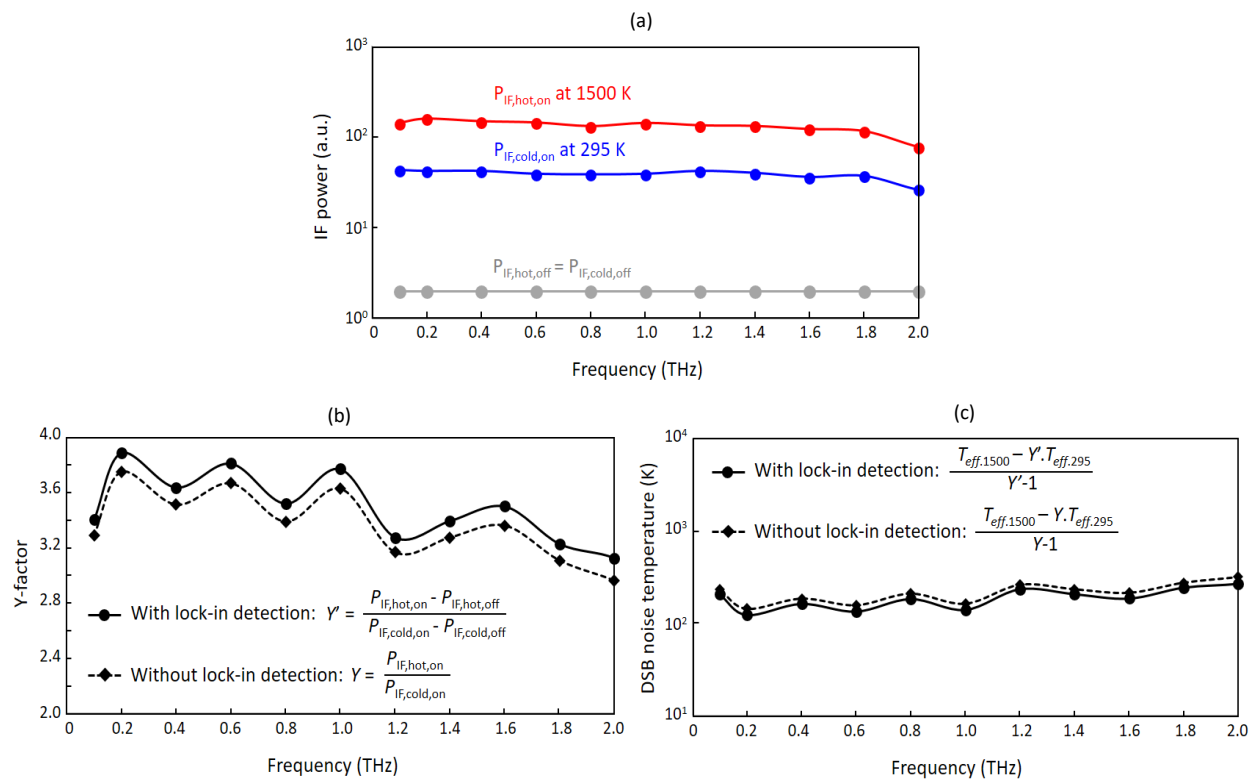

Figure 2. (a) The measured values of $P_{IF,hot,on}$, $P_{IF,cold,on}$, $P_{IF,hot,off}$, and $P_{IF,cold,off}$ for hot and cold temperatures of 1500 K and 295 K, respectively. The calculated Y-factor and DSB noise temperature values with and without lock-in detection are shown in (b) and (c), respectively. The differences between the Y-factor and DSB noise temperature values calculated with and without lock-in detection are lower than the fluctuations in the Y-factor and DSB noise temperature values due to the fluctuations in the optical pump power (error bars shown in Fig. 3a and Fig. 3c of the manuscript).

D. Conversion Gain

Zmuidzinas et al. have made yet another inaccurate claim about measuring the conversion loss/gain of our photomixer:

'*The paper provides no information on the mixer conversion loss, an important quantity that could readily have been measured and reported as a consistency check. The paper thus offers no reliable experimental evidence that substantiates the claimed sensitivities.*'

It is true that the overall THz-to-RF conversion loss/gain of the heterodyne detection system can be measured, however, extracting the mixer conversion loss/gain requires knowledge of all of the loss values in the system including, atmospheric propagation loss, lens reflection/absorption loss, antenna coupling loss, IF coupling loss, and the noise temperature of the IF chain, some of which cannot be directly measured. Therefore, an estimate of these loss parameters can be used to get an estimate of the mixer conversion loss/gain, not a measurement of the mixer conversion loss/gain.

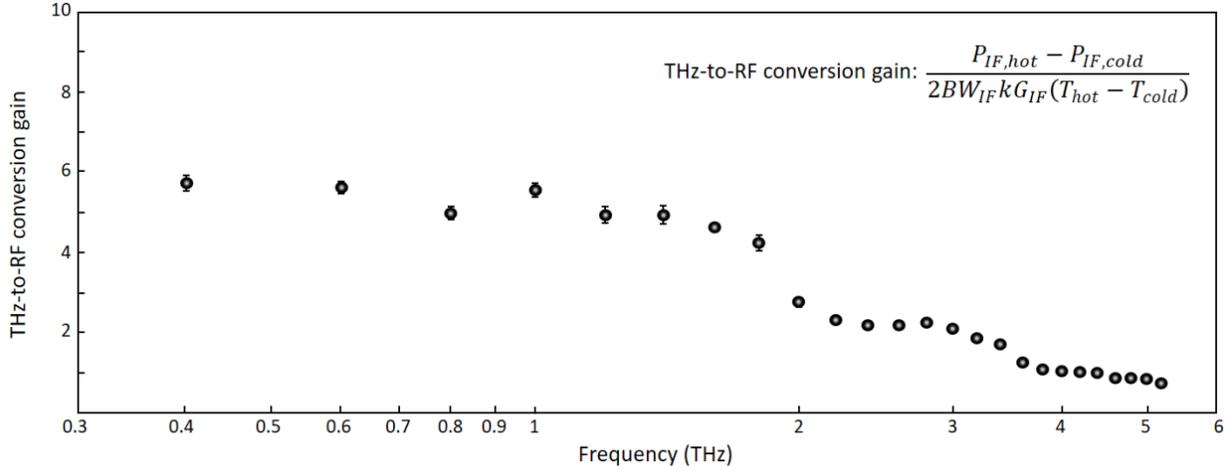

Figure 3. The extracted terahertz-to-RF conversion gain of the heterodyne detection system based on plasmonic photomixing from the Y-factor measurements.

Figure 3 shows the extracted terahertz-to-RF conversion gain of our heterodyne detection system from the Y-factor measurements using

$$\frac{P_{IF,hot} - P_{IF,cold}}{2BW_{IF}.k.G_{IF}(T_{hot} - T_{cold})}$$

where $P_{IF,hot}$ and $P_{IF,cold}$ are the IF output power in response to hot and cold loads, $G_{IF}$ is the gain of the IF chain, $k$ is Boltzmann's constant, and $BW_{IF}$ is the bandwidth of the IF bandpass filter. Unlike the incorrect claim of Zmuidzinas et al. that says '*Internal conversion gain in this device seems implausible, especially such a large value, given the lack of any measurements of conversion loss and the absence of a clear physical mechanism for gain*', the measured high conversion gain values are also theoretically predicted given the large density of the photocarriers in close proximity of the plasmonic contact electrodes of the photomixer, which oscillate at a terahertz beat frequency and drift to the contact electrodes in response to the received terahertz electric field, which is also enhanced in close proximity to the contact electrodes. The detailed derivations of the device governing equations and the numerical simulations of these enhancement factors are described in the supplementary Figs S2 and S3.

Zmuidzinas et al. have also questioned the noise temperature of our IF system:

'*Wang et al. also do not report measurements for $T_{IFsystem}$ or TLNA, although according to the manufacturer's data sheet for their Mini Circuits ZRL-1150 first-stage amplifier, we may take TLNA ≥ 70 K as a reasonable value including cable losses.*'

The measured noise temperature of the IF chain including the LNA, BPF, and cables is 63.5 K. This value does not include the noise temperature contributions of the power meter and lock-in amplifier. Noise figure meters cannot measure the correct noise figure of a power meter because it measures the incident power from the other components in the IF chain plus the injected noise. Neither they can measure the correct noise figure of the lock-in amplifier because their noise source cannot be modulated as fast of the 100 kHz modulation rate used in our measurements.

E. Heterodyne Operation

Zmuidzinas et al. provided a suggestion to demonstrate the heterodyne operation of our detector based on plasmonic photomixing:

*'Experimentally demonstrate that the receiver response is truly heterodyne, e.g., by using infrared-blocking filters, THz passband filters, and ideally gas-cell measurements of molecular absorption lines.'*

We have already performed gas cell measurements of the molecular absorption lines of ammonia using our heterodyne terahertz detector based on plasmonic photomixing and resolved the spectral signatures of ammonia over a 1-5 THz frequency range, as described in reference [3].

The use of terahertz filters is essential in most heterodyne terahertz detection systems based on mixers that are capable of direct terahertz detection, e.g., diode and HEB mixers, to eliminate the impact of the background IR radiation. However, the heterodyne detection system based on photomixing does not require such filtering because it is not capable of direct terahertz detection. To experimentally prove that the background IR radiation does not impact our Y-factor measurements, we have turned off one of the pump lasers to switch the optical pump beam from heterodyning to continuous wave. Under the same optical pump power, the photomixer output does not vary when switching between the hot and cold loads, i.e. the Y-factor values fall below noise, confirming that the heterodyne detection process is not impacted by the background IR radiation.

In summary, all of the results and conclusions reported in our *Nature Astronomy* manuscript, DOI: 10.1038/s41550-019-0828-6, remain correct, as before.